\documentclass[11pt,twoside]{article} 
\usepackage{asp2004}
\usepackage{epsf}
\usepackage{psfig}
\usepackage{lscape} 

\begin{document} 
\title{Disentangling Composite Spectrum Hot Subdwarfs}
\author{M.\ A.\ Stark and Richard A.\ Wade} 
\affil{The Pennsylvania State University\\525 Davey Lab, University Park,
 PA 16802 USA}
\begin{abstract} 
We give preliminary results from a spectroscopic study of composite
spectrum hot subdwarfs (sd+late-type).  We obtained spectra of a
sample of hot subdwarfs selected from the {\em Catalogue of
Spectroscopically Identified Hot Subdwarfs} {\citep*{KHD}} on the
basis of near-infrared photometry from the {\em Two Micron All Sky
Survey} (2MASS).  The sample consists of 20 photometric and
spectroscopic single and 54 composite hot subdwarfs, 6 resolved (or
barely resolved) visual doubles, and 5 objects with emission lines or
broad absorption lines with emission cores.  Spectra of 84
``standard'' (single late-type) stars with {\em Hipparcos} parallaxes
were also obtained for calibration.  These observations cover
4600--8900~{\AA} with $\sim$3~{\AA} resolution.  We measured
equivalent width-like indices around Mg~I~b, Na~I~D, the Ca~II
infrared triplet, H$\alpha$, and H$\beta$.  Using the single late-type
star observations combined with model energy distributions, we explore
how the measured indices of a composite spectrum vary as the
temperature and luminosity of the late-type companion are varied and
as the temperature and radius of the hot subdwarf are varied.  We use
the measured indices of the composite systems to estimate the
temperature and gravity of the late-type star, taking into account the
dilution of its spectral features by light from the hot subdwarf.
\end{abstract}

\section{Samples and Observations}
Hot subdwarfs were selected from the {\emph{Catalogue of
Spectroscopically Identified Hot Subdwarfs}} {\citep{KHD}}.  The
observed sample consists of 20 single and 54 composite hot subdwarfs
(primarily classified as sdB) that have $V \la 13.5$.  Composite
subdwarfs for this sample were defined to have $J\!-\!K_S \ga$~$+0.05$
{\citep{SW}}.  There are also 6 subdwarfs that have close ($<$10$''$)
resolved companions, and 5 that show emission lines (4 met our
composite color criterion).

Eighty-four stars (52 main sequence, 32 subgiant) with parallax
measurements from {\emph{Hipparcos}} were selected to provide
empirical calibration.  They cover the Pop~I main sequence from early
F to early M-type, and post-main sequence subgiants from F to K-type
($2.7\la V\la 10.0$, $-1.1$ $\la \mathrm{M}_V \la 9.2$, and $0.35\la
\bv\la 1.6$).

\begin{figure}[!ht]
\resizebox{\textwidth}{!}{\rotatebox{-90}{\plotone{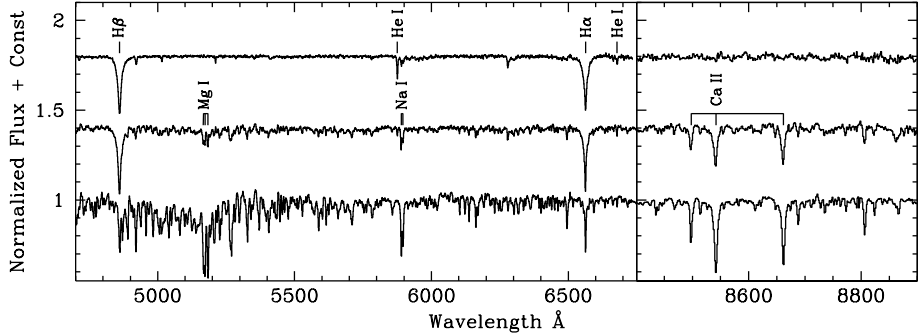}}}
 \caption{Example spectra for a single hot subdwarf (top), a composite
 hot subdwarf (middle), and a K1V standard (bottom) offset by
 constants.  It is evident that there is less dilution at longer
 wavelengths. \label{Spec}}
\end{figure}

Spectra were taken with the GoldCam spectrograph on the Kitt Peak
National Observatory 2.1m telescope.  The wavelength interval
4600--8900~{\AA} was covered at $\sim$1.3~{\AA/pixel}, using two
spectrograph setups.  This wavelength range includes H$\alpha$,
H$\beta$, H-Paschen 11$\rightarrow$limit, Mg~I~b, Na~I~D, Ca~II
infrared triplet (CaT), {He~I} 5875 and 6678~{\AA}.\@ Unfortunate
observing conditions prevented accurate flux calibration on most
nights, so we consider the normalized spectra only.  Example spectra
are shown in Fig.~\ref{Spec}.

\section{Equivalent Width Measurements \label{EW}}
Equivalent width (EW) indices were measured for prominent lines after
fitting local continua in the regions of the lines of interest.  We
focus on the {CaT}, {Na~I}~D, and {Mg~I}~b lines.

Kurucz flux distributions were matched to the cool standards based on
{\bv} colors, with $\log g=4.5$ used for main sequence standards and
lower gravities for the subgiants.  High gravity ($\log g=5.0$) models
with {T$_{\mathrm{eff}}$}~= 20000, 26000, and 32000~K were used as
proxies for hot subdwarfs.  The observed EW for the measured lines of
each of the standard stars were diluted by creating a composite with
one of the three proxies, using the flux distributions to scale the
contributions of the two stars appropriately.  For a given hot
subdwarf+cool companion combination, each of the EW indices is diluted
by a different amount, according to the wavelength-dependent ratio of
fluxes from the two stars, assuming radii for the hot subdwarfs from
{\citet[][zero age horizontal branch]{Caloi}} and main sequence radii
from {\citet{Gray}} (subgiants were scaled based on their {\em
Hipparcos} {M$_V$}).

\section{Equivalent Width Comparison \label{EWcompare}}
\begin{figure}[!ht]
\plottwo{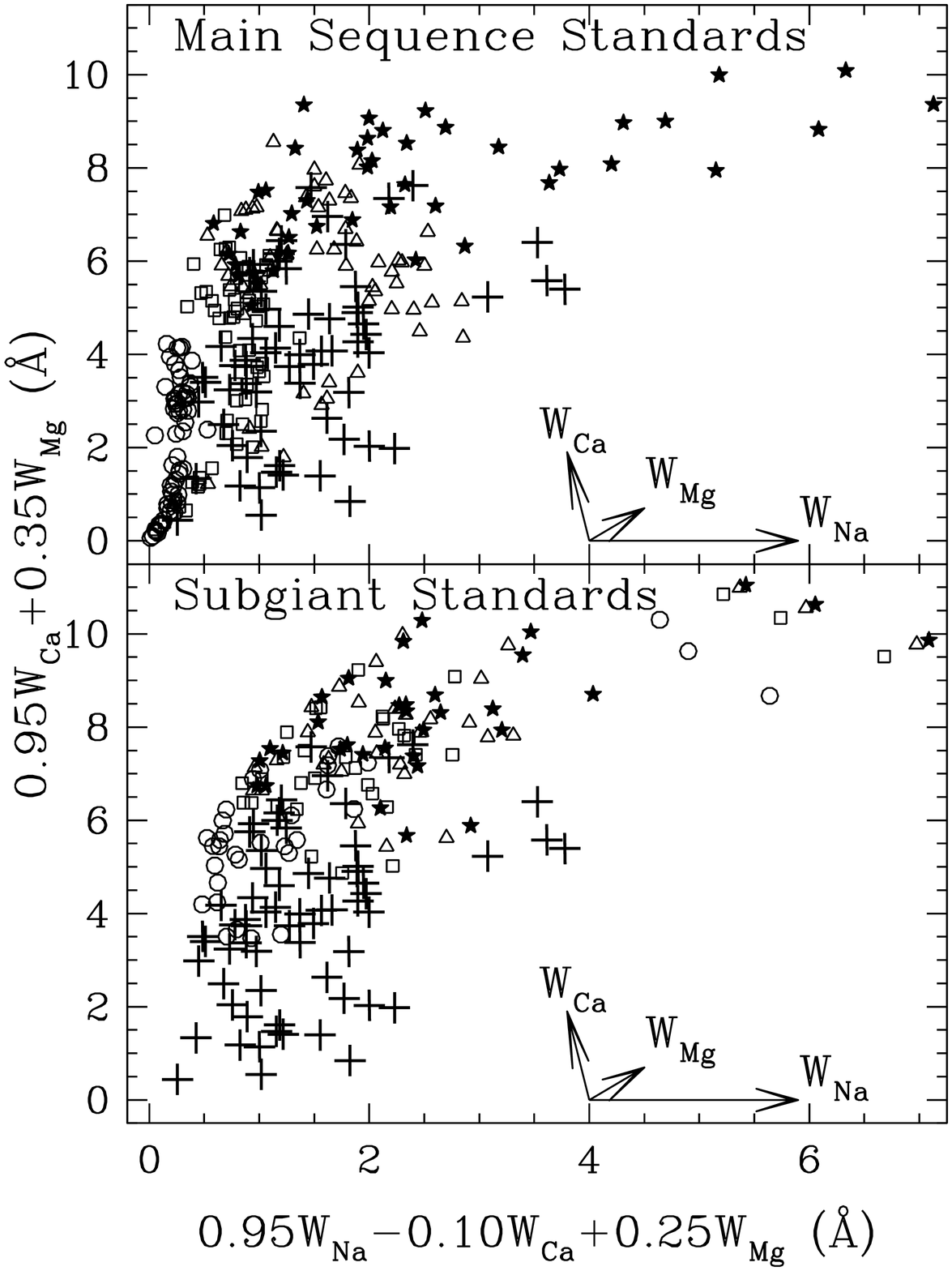}{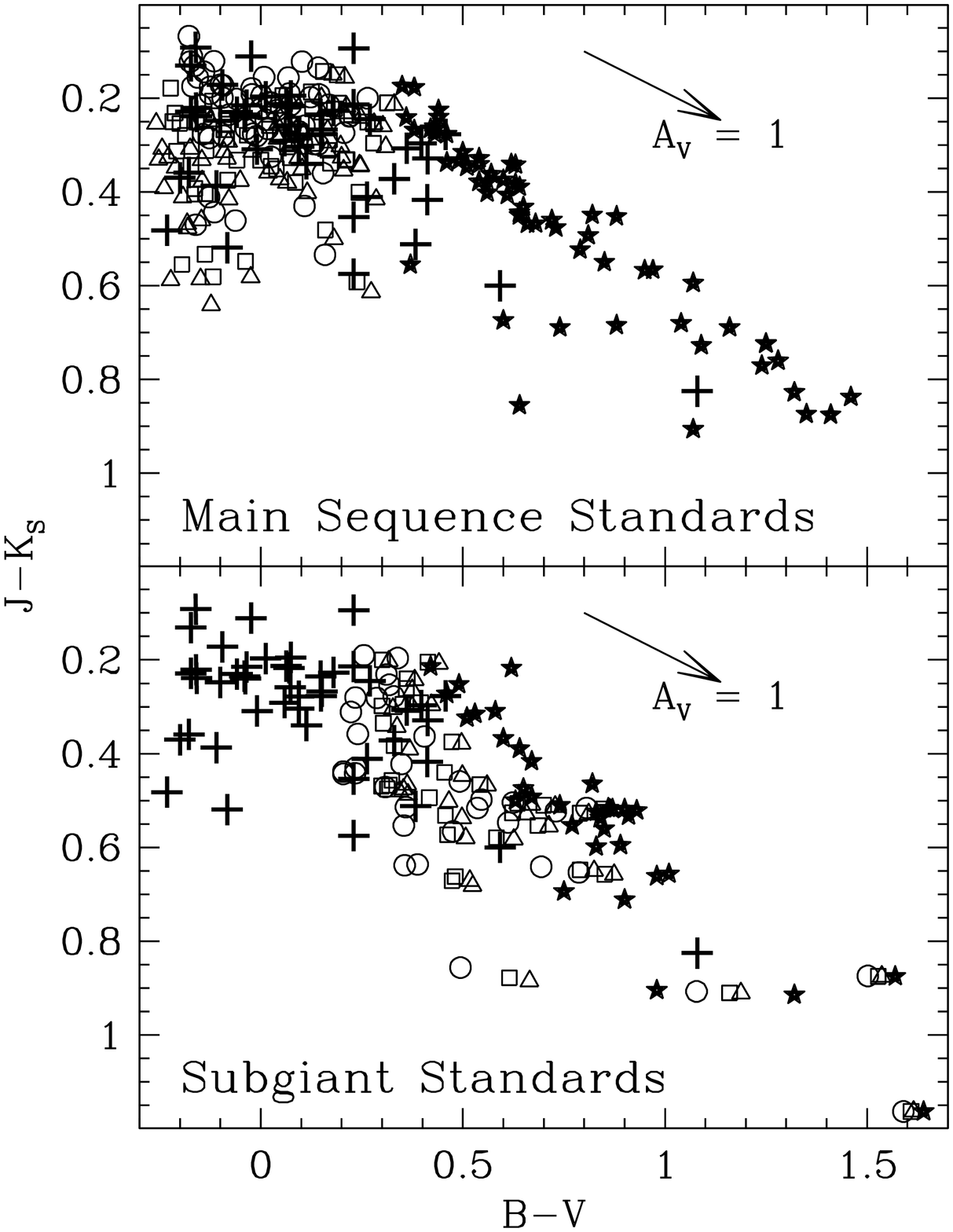}
 \caption{Panel $a$ (left): EWs for diluted main sequence stars (top)
 and subgiants (bottom) compared with observed composite hot
 subdwarfs.  Panel $b$ (right): Color indices for diluted main
 sequence stars (top) and subgiants (bottom) compared with observed
 composite hot subdwarfs. Both panels: +~= observed composite sd;
 star~= undiluted standards; circle~= model EW or colors, each
 standard diluted by a 20kK hot subdwarf; square~= 26kK dilution;
 triangle~= 32kK dilution. \label{Ratio}}
\end{figure}

The three EWs ({CaT}, {Na~I}, {Mg~I}) form a three dimensional space.
Most of the information can be shown in the 2-D subspace defined by
([$0.95 W_{\mathrm{Na}}$ $-0.10 W_\mathrm{Ca}+0.25 W_\mathrm{Mg}$],
[$0.95 W_\mathrm{Ca}+0.35 W_\mathrm{Mg}$]).  Using this projection,
EWs for composite hot subdwarfs are plotted in {Fig.~\ref{Ratio}}$a$
with diluted and undiluted EWs from main sequence standards (upper
panel), and subgiant standards (lower).  Three arrows indicate the
direction and magnitude of $\Delta W=2$~{\AA} for {CaT}, {Mg~I}, and
Na~I.\@

In both panels of {Fig.~\ref{Ratio}}$a$, cooler {\em undiluted}
standard stars (star symbols) are to the right, hotter to the
left.  However, in the upper panel, the cooler (fainter) {\em diluted}
main sequence standards are at the bottom, while in the lower panel the
cooler (brighter) {\em diluted} subgiants are to the upper right (in
both cases the 20kK hot subdwarf causes the greatest amount of
dilution, while the 32kK subdwarf causes the least).  Both effects are
due to the respective trends of {M$_{V}$} vs.\ {T$_\mathrm{eff}$}.

From {Fig.~\ref{Ratio}}$a$, it is seen that the EW indices can in most
cases be accounted for by composite models involving main sequence
companions; models with subgiants are less successful.

\section{Diluted Color Indices \label{DilutedCI}}
Diluted color indices were calculated using observed visual and 2MASS
magnitudes for the standards, and combining them with theoretical
colors for the three hot subdwarf temperatures (32kK, 26kK, 20kK)
based on {\citet{Girardi}} with 2MASS color transformations taken from
{\cite{Carpenter}}.  The contributions from the two stars were scaled
using {M$_V$} from {\em Hipparcos} for the standards,
$L_{\mathrm{bol}}$ for the hot subdwarfs from {\citet{Caloi}}, and
bolometric corrections for both standards and hot subdwarfs from
{\citet{Girardi}}.  Standards were matched to models based on {\bv}
color.  The color indices (\bv, {$J\!-\!K_S$}) are shown for the
observed composite hot subdwarfs, and for each standard diluted by a
32kK, a 26kK, and a 20kK hot subdwarf in Fig.~\ref{Ratio}$b$; in this
color space composite-colored stars are clearly separated from
single-colored stars {\citep[see also][]{SW}}.

As with the EWs (\S\ref{EWcompare}), the color of most composite hot
subdwarfs can be explained using models involving main sequence
companions, while models with subgiants are less successful.

\section{Individual Models \label{Fit}}
Treating the diluted EW measurements and diluted 2MASS colors indices
for each standard star as a ``model'', the $\chi^2$ value can be
calculated between each composite subdwarf and each ``model'' to find
a best fit.  Preliminary results indicate that most composite hot
subdwarfs in our sample favor 26kK or 32kK hot subdwarfs with {\em
main sequence companions} in the range of $0.4 \la \bv \la 1$ (roughly
late-F to mid-K spectral type).  However, there are a few objects that
favor only subgiant models, and a few that equally accept a hotter
(26kK or 32kK) hot subdwarf with a main sequence companion, or a
cooler (20kK) hot subdwarf with a subgiant companion.  More work is
needed to break these degeneracies; in particular temperature
estimates (from UV spectra or GALEX observations) for the hot
subdwarfs would be useful.  Also, we will replace the sparse and noisy
observational model grid with a smooth grid interpolated from the
observed behaviors.

\section{Summary \label{Summary}}
We obtained spectra of composite colored hot subdwarfs and single cool
standard stars with {\emph{Hipparcos}} parallaxes.  Preliminary
analysis indicates {\em the majority of cool companions in composite
hot subdwarf systems are consistent with the main sequence}.  Overall,
subgiant companions do not explain the observed colors and EWs for the
majority of composite hot subdwarfs; but they are not excluded in some
cases if the hot subdwarf is relatively cool (i.e., $\sim$20kK).
Further analysis of additional features in individual spectra and
refinement of the model grids is needed to distinguish between main
sequence and subgiant in these cases.  An independent measure of the
temperatures of the hot subdwarfs (from IUE spectra or GALEX fluxes)
would further constrain uncertain cases.

 \acknowledgements{We thank T.~Bogdanovic, J.~Ding, K.~Herrmann,
 \linebreak K.~Lewis, and A.~Narayanan for assisting with the
 observations.  This publication makes use of data products from the
 Two Micron All Sky Survey, a project of the University of
 Massachusetts and the Infrared Processing and Analysis
 Center/California Institute of Technology.  This research has been
 supported in part by: NASA grant NAG5-9586, NASA GSRP grant
 NGT5-50399, the Zaccheus Daniel Foundation for Astronomical Science,
 Sigma Xi Grant-in-Aid of Research, NOAO graduate student travel
 support, and a NASA Space Grant Fellowship through the Pennsylvania
 Space Grant Consortium.}

\end{document}